\documentclass[
prmaterials,
reprint,               
superscriptaddress,     
floatfix
]{revtex4-2}

\usepackage[T1]{fontenc}
\usepackage[utf8]{inputenc}
\usepackage{lmodern}
\usepackage{amsmath,amssymb,amsfonts}
\usepackage{graphicx}
\usepackage{xcolor}
\usepackage{microtype}
\usepackage{hyperref}
\hypersetup{
	colorlinks=true,
	linkcolor=blue,
	citecolor=blue,
	urlcolor=blue
}

\begin{document}
	
	\title{Tuning the optoelectronic and magnetic properties of Penta-PtN$_2$ nanoribbons via edge engineering and defects}

	\author{Le Thi Thuy My}
	\affiliation{College of Natural Sciences, Can Tho University, 3-2 Road, Can Tho City 94000, Vietnam} 
	\affiliation{Faculty of Electrical and Electronics , Vinh Long University of Technology and Education, Vinh Long City 85100, Vietnam}
	
	\author{Pham Thi Bich Thao}
	\affiliation{College of Natural Sciences, Can Tho University, 3-2 Road, Can Tho City 94000, Vietnam}
	
	\author{Nguyen Hai Dang}
	\affiliation{Faculty of basic, Nam Can Tho University, Can Tho City 900000, Vietnam}
	
	\author{Nguyen Thanh Tien}
	\email{nttien@ctu.edu.vn}
	\affiliation{College of Natural Sciences, Can Tho University, 3-2 Road, Can Tho City 94000, Vietnam}
	
	\date{\today}

\begin{abstract}
In this study, we investigate aspects including the structural, electronic, optical and magnetic properties of the p-PtN$_2$ nanoribbons with four typical edge shapes of sawtooth–sawtooth (SS), armchair–armchair (AA), zigzag–armchair (ZA), and zigzag–zigzag (ZZ). Along with this, based on density functional theory (DFT) calculations, we obtained the values of important quantities including the binding energies, spin-splitting electronic band structures, atom-projected density of states (PDOS), spatial spin density distribution, optical absorption spectra, magnetic properties and investigation of vacancy defects including single Pt vacancy, single N vacancy, Pt–N double vacancy, and N divacancy.The results demonstrate that all the p-PtN$_2$NR structures are found to have good structural stability. For the electronic properties, AA-7, AA-9, AA-11, and AA-13 behave as ferromagnetic semiconductors in both spin channels, SS-11 shows ferromagnetic half-metal, and the other configurations are metallic. The optical absorption spectra of the penta PtN$_{2}$ nanoribbon  indicates that the absorption spectral region can be flexibly controlled by varying the ribbon width and edge geometry. The loss of one or several atoms within the structure also alters the electronic properties, as evidenced by the ZZ7 configuration; specifically, the defective structure exhibits semi-metallic behavior instead of the metallic nature observed in the pristine sample. Additionally, the defective ZZ7 samples shift their optical absorption range from the near-infrared to the visible light region.

\end{abstract}

\maketitle

\section{Introduction}\label{sec:intro}
A new era in the history of nanomaterial research began in 2004, Geim and Novoselov were the first scientists to successfully isolate monolayer graphene from graphite \cite{novoselov2004electric}. This discovery marked the significance of two-dimensional (2D) materials in materials research. Building on this success, The scientific community has expanded the scope of research to other 2D material systems through both experimental and theoretical approaches, including silicene, germanene, and phosphorene \cite{guzman2007electronic,vogt2012silicene}. However, A major challenge arises because of most of the discovered 2D structures possess hexagonal lattices with either excessively large band gaps (h-BN) \cite{watanabe2004direct} or gapless (graphene) \cite{castro2009electronic} which significantly constrains its applicability in optoelectronic devices.

To address this issue, in $2015$, Zhang et al. first proposed penta-graphene, a two-dimensional material composed entirely of pentagonal rings (a new carbon allotrope composed entirely of pentagonal rings), thereby inaugurating the family of pentagonal two-dimensional materials (penta-materials) \cite{zhang2015penta}. Generally, these pentagonal materials can be divided into four subgroups based on their constituent elements: unary (monoatomic/elemental), binary, ternary, and quaternary \cite{shah2024theoretical}. Since the emergence of penta-graphene has stimulated a powerful wave of research, leading to the reporting of a series of elemental pentagonal materials, such as penta-silicene \cite{ding2015hydrogen} and subsequently expanding to binary and ternary pentagonal systems, including penta-NiN$ _{2} $ \cite{bykov2021realization}, penta-PdSe$ _{2}$ \cite{lei2019new}, penta-PtN$ _{2}$, Penta-BCN \cite{zhao2020penta}, Penta-CNP \cite{sun20211}, penta-BNSi \cite{varjovi2022ternary}.Two-dimensional materials with pentagonal structures are considered promising candidates for nanoelectronic devices owing to their unique structures and electronic, mechanical, optical, and other exceptional properties \cite{han2024band}. 
Recently, transition metal nitrides have attracted considerable attention owing to their outstanding properties in terms of mechanical strength, toughness, and remarkable optical, electronic, and magnetic characteristics \cite{crowhurst2006synthesis}. Platinum pernitride (PtN$ _{2}$) is a transition-metal nitride characterized by the presence of N–N bonds within its crystal lattice. This compound was first discovered in the Pt–N system under ultra-high pressure conditions ($>$ 50 GPa) using a laser-heated diamond anvil cell (LH-DAC) technique. PtN$ _{2}$ crystallizes in a pyrite-type structure and has been confirmed to be dynamically stable under ambient conditions, although it is formed only at very high pressures. However, because the PtN$ _{2}$ crystals obtained by conventional LH-DAC methods are extremely small, experimental investigations of their electronic and optical properties have been significantly limited. To overcome this issue, Niwa et al. developed a nitridation method of Pt thin films deposited on an $\alpha$-Al$ _{2}$O$ _{3}$ substrate under pressures of approximately 50 GPa, thereby successfully synthesizing large-area polycrystalline PtN$ _{2}$ thin films with high structural quality. This approach not only enables reliable electrical and optical characterizations but also experimentally demonstrates that PtN$ _{2}$ is a semiconductor with a bandgap of about 2 eV, in good agreement with previous first-principles predictions \cite{niwa2022high}. Along with the rapid development of two-dimensional materials science, extending PtN$ _{2}$ to lower-dimensional limits has become a promising research direction. In this context, a two-dimensional penta-PtN$ _{2}$ phase with a pentagonal lattice has been proposed as a special derivative of bulk PtN$ _{2}$.

Theoretical studies on two-dimensional pentagonal PtN$ _{2} $ have emerged, showing that this material exhibits many outstanding properties. In $2019$, using the particle swarm optimization (PSO)-based global structure search method combined with first-principles calculations, Zhao et al. identified a new family of two-dimensional planar penta-MN$ _{2}$ (M = Pd, Pt) monolayers. These metal dinitride sheets, composed entirely of pentagons and containing N$ _{2}$ dimers, exhibit outstanding dynamical, thermal, and mechanical stability, while being energetically more favorable than the experimentally synthesized pyrite MN$ _{2}$ phases. In particular, penta-PtN$ _{2}$ is a direct band-gap semiconductor (75 meV) with ultrahigh carrier mobility comparable to that of graphene and can withstand temperatures above 2000 K, making it a highly promising candidate for high-temperature nanoelectronic and optoelectronic applications \cite{zhao20192d}. In addition, one study has demonstrated that the electronic and optical properties of penta-PtN$ _{2}$ can be effectively tuned via strain engineering. In particular, compressive strain can induce a semiconductor-to-metal phase transition, significantly enhancing carrier transport and markedly improving absorption in the visible-light region, thereby enabling tunable optoelectronic responses and device switching behavior. This pronounced strain-dependent tunability highlights the high flexibility of penta-PtN$ _{2}$ in device design \cite{han2024band}. Another study have demonstrated that penta-PtN$ _{2}$ is a potential pentagonal anode material for alkali-metal ion batteries, featuring low alkali-ion migration barriers, moderate capacity, and stable electrochemical behavior at high temperatures\cite{chen2021modelling}.

The outstanding properties of penta-PtN$ _{2}$ reported in previous studies have motivated us to investigate the one-dimensional penta-PtN$ _{2}$ nanoribbon structures, is highly necessary. Investigating the one-dimensional (1D) structures of materials can give rise to unique physical properties, enhanced structural flexibility, and superior application potential, for which the 1D structures of penta-graphene serve as a compelling example \cite{yuan2017electronic}. Most recently, the study of the electronic, magnetic, and optical properties of $p$-$NiN_2$ nanoribbons by Dang et al. has further clarified the unique properties observed in 1D structures \cite{dang2025diverse}. For these reasons, in this work, we study the stability structure, electronic properties, optical, and magnetic properties of four basic edge shapes with widths ranging from 5 to 15 of penta PtN$ _{2} $ nanoribbons (p-PtN$ _{2} $NRs) were selected to  phenomena using density functional theory. At the same time, in this study, we investigate the structural stability as well as the electronic, magnetic, and optical properties of several defective models in order to explore properties that are superior to those of the pristine structure.

\section{Computational details}
The p-PtN$_2$ nanoribbon structures are cut from the 2D p-PtN$_2$ along different directions. The resulting 4 possible types of nanoribbons, including sawtooth–sawtooth (SS), armchair–armchair (AA), zigzag–armchair (ZA), and zigzag–zigzag (ZZ). Hydrogen atoms are employed to passivate the edges of the four nanoribbon configurations in order to eliminate dangling bonds. The structural stability of AA, SS, ZA, and ZZ nanoribbon structures with widths varying from 5 to 15 is evaluated based on the binding energy calculated using the following equation.
\begin{equation}\label{key}
	E_B =\frac{E_{\texttt{total}} - n_{\texttt{Pt}}E_{\texttt{Pt}} - n_{\texttt{N}}E_{\texttt{N}} -  n_{\texttt{H}}E_{\texttt{H}}}{n_{\texttt{Pt}} + n_{\texttt{N}} + n_{\texttt{H}}}
\end{equation}

In equation (1), the binding energy of studied structures, the ground-state energy of the total system, the ground-state energy of isolated Pt, N, and H systems are denoted $E_B$ (eV/atom), $E_{\texttt{total}}$, $E_{\texttt{Pt,N,H}}$ (eV), respectively. The numbers of Platinum, Nitrogen and Hydrogen atoms in the unit cell are described by $n_{\texttt{Pt,N,H}}$.

In this work, the HSE06 method implemented in Atomistix ToolKit (ATK) \cite{Stokbro2010} is employed to improve the accuracy of bandgap and optical property calculations . The calculations utilized PseudoDojo pseudopotentials and a linear combination of atomic orbitals (LCAO) basis set \cite{Smidstrup2019}. TThe one-dimensional ribbon models are constructed to be periodic along the z-direction, with confinement in the x and y directions. A vacuum layer of 15 Å is introduced to suppress interactions between neighboring images. The density mesh cutoff is maintained at 2000 eV throughout geometry optimization and self-consistent calculations. Meanwhile, the k-points set at 1 $\times$ 1 $\times$ 70 and 1 $\times$ 1 $\times$ 200 for the optimal and self-consistent calculations, respectively, while the 0.001 eV/Å and $10^{-6}$ eV are the convergence values of the force and energy set during calculations.

The optical absorption coefficient is obtained from \cite{quinten2010optical}
\begin{equation}\label{eq2}
	\alpha  = \frac{{2\omega \kappa }}{c},
\end{equation}
where, $c$, $\kappa$ respectively denote the speed of light and the extinction coefficient.
\begin{equation}\label{eq3}
	\kappa  = \sqrt {\frac{{\sqrt {\varepsilon _1^2 + \varepsilon _2^2} }}{2} - \frac{{{\varepsilon _1}}}{2}}.
\end{equation}

$\varepsilon _1$ and  $\varepsilon _2$ are the real and imaginary parts of the dielectric function
\begin{equation}\label{eq4}
	\varepsilon \left( \omega  \right) = {\varepsilon _1}\left( \omega  \right) + i{\varepsilon _2}\left( \omega  \right),
\end{equation} 
$\omega$ is the angular frequency of the incident photon.

\section{Results and discussions}

\subsection{Structural properties}

\begin{figure*}[h]
	\begin{center}
		\includegraphics[angle=0, width=0.8\textwidth]{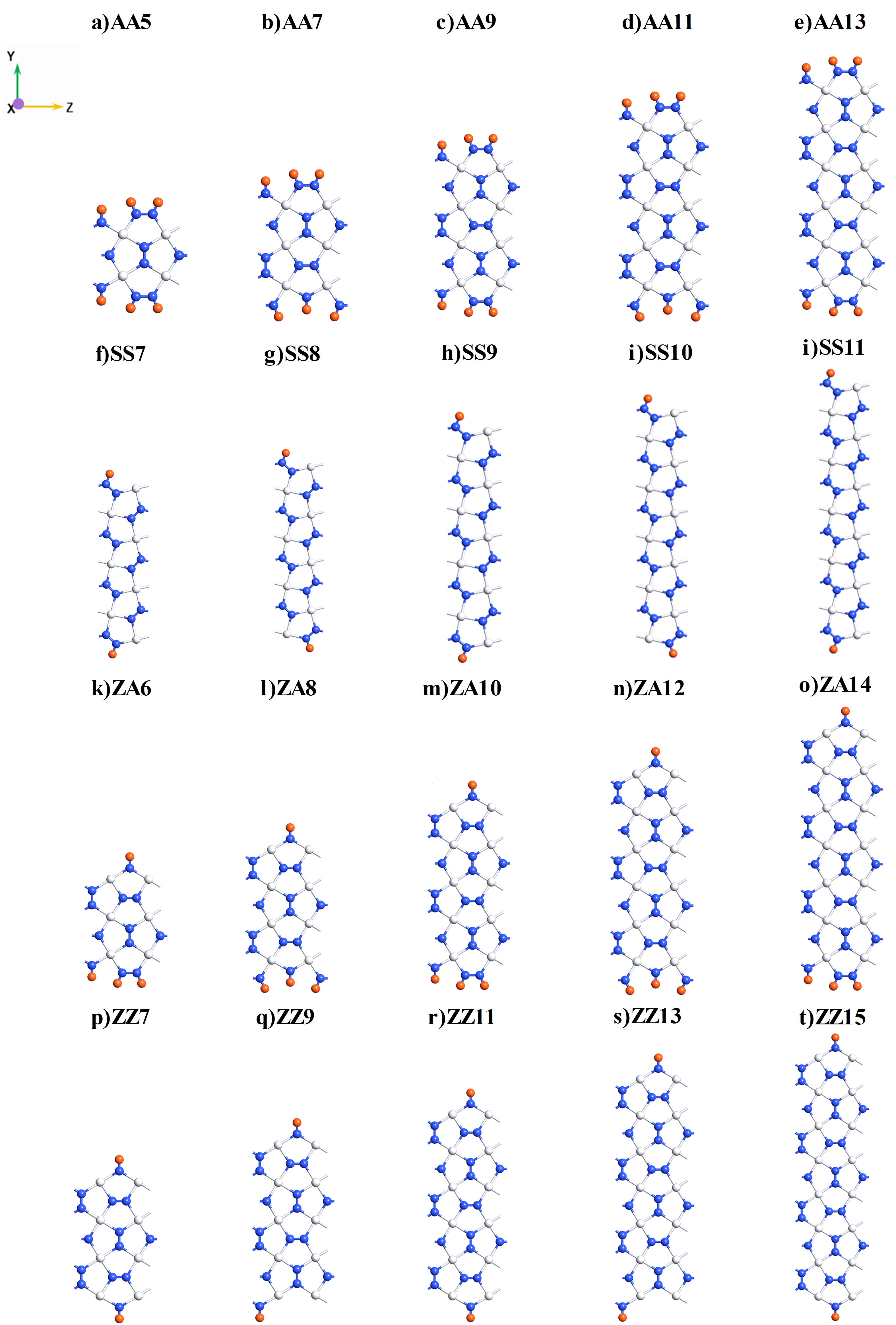}
		\caption{ Top-view model of the p-PtN$_2$NRs with 20 optimal configurations for four edge types. The white, blue, and orange balls represent Pt, N and H atoms, respectively.}
		\label{Fig1} 
	\end{center} 
\end{figure*}

After optimization, the geometric configurations of the p-PtN$_2$ nanoribbons (p-PtN$_2$NRs), presented in top-view projection with twenty different configurations having H-passivated edges, are shown in Fig. \ref{Fig1} The widths of the studied configurations range from 5 to 15 dimer lines along the y-direction. These optimized configurations are subsequently utilized to investigate their electronic, optical, and electronic transport properties.

\begin{figure}[h]
	\begin{center}
		\includegraphics[angle=0, width=0.5\textwidth]{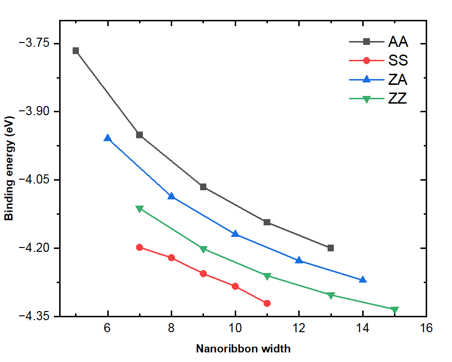}
		\caption{Formation energy plotted as a function of the nanoribbon width for the four edge types of p-PtN$_2$NRs.}
		\label{Fig2} 
	\end{center} 
\end{figure}

The structural stability is evaluated through the binding energy of each configuration. Accordingly, the binding energies of the configurations investigated in this study were calculated and are presented in Fig. \ref{Fig2}, in which all binding energy values are significantly negative, in the range of $-3.8$ to $-4.3$ eV. These significant negative values indicate the good stability of p-PtN$_2$NRs in Fig. \ref{Fig2}, the binding energies of the AA, SS, ZA, and ZZ structures are represented by the red, magenta, blue, and black lines, respectively. As observed in Fig. \ref{Fig2}, the binding energies of all edge configurations exhibit a similar trend: the binding energy decreases with increasing ribbon width. Since a lower binding energy indicates higher structural stability, nanoribbons with larger widths are therefore more stable.
Figure. \ref{Fig2} clearly demonstrates that the SS structure possesses the lowest binding energy among the four p-PtN$_2$ nanoribbon configurations, indicating that the SS-p-PtN$_2$ structure is the most stable. The remaining structures, arranged in ascending order of binding energy, are ZZ-p-PtN$_2$, ZA-p-PtN$_2$, and AA-p-PtN$_2$. This behavior is in excellent agreement with the results previously reported for penta nanoribbon structures\cite{yuan2017electronic, mi2021diverse,dang2025diverse}.

\subsection{Electronic and optical properties}

\begin{figure*}[h]
	\begin{center}
		\includegraphics[angle=0, width=0.7\textwidth]{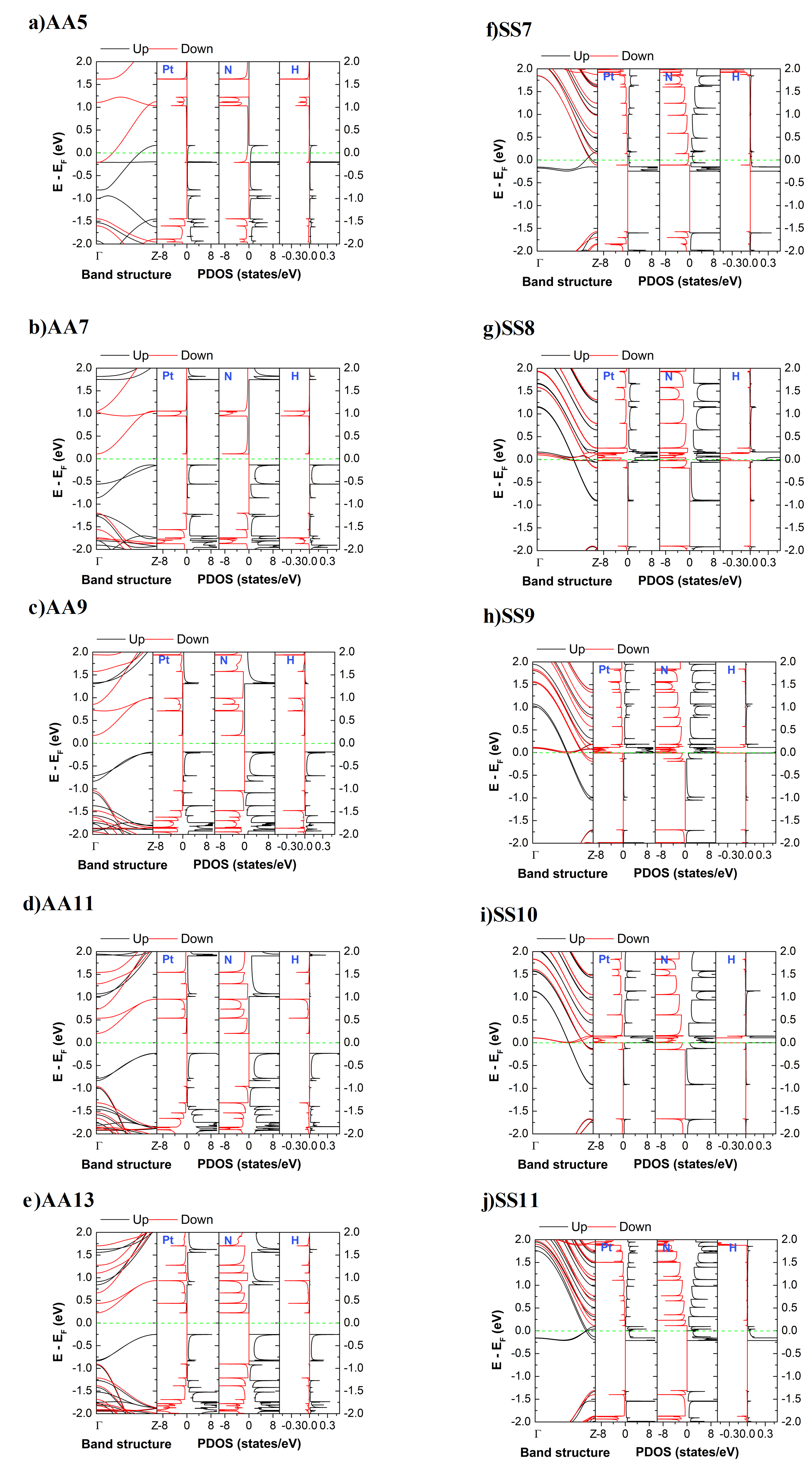}
		\caption{Spin-splitting band structure and atom-projected density of states (PDOS) for the AA and SS edge shapes of the p-PtN$_2$NRs with differrent widths.}
		\label{Fig3} 
	\end{center} 
\end{figure*}
\begin{figure*}[h]
	\begin{center}
		\includegraphics[angle=0, width=0.7\textwidth]{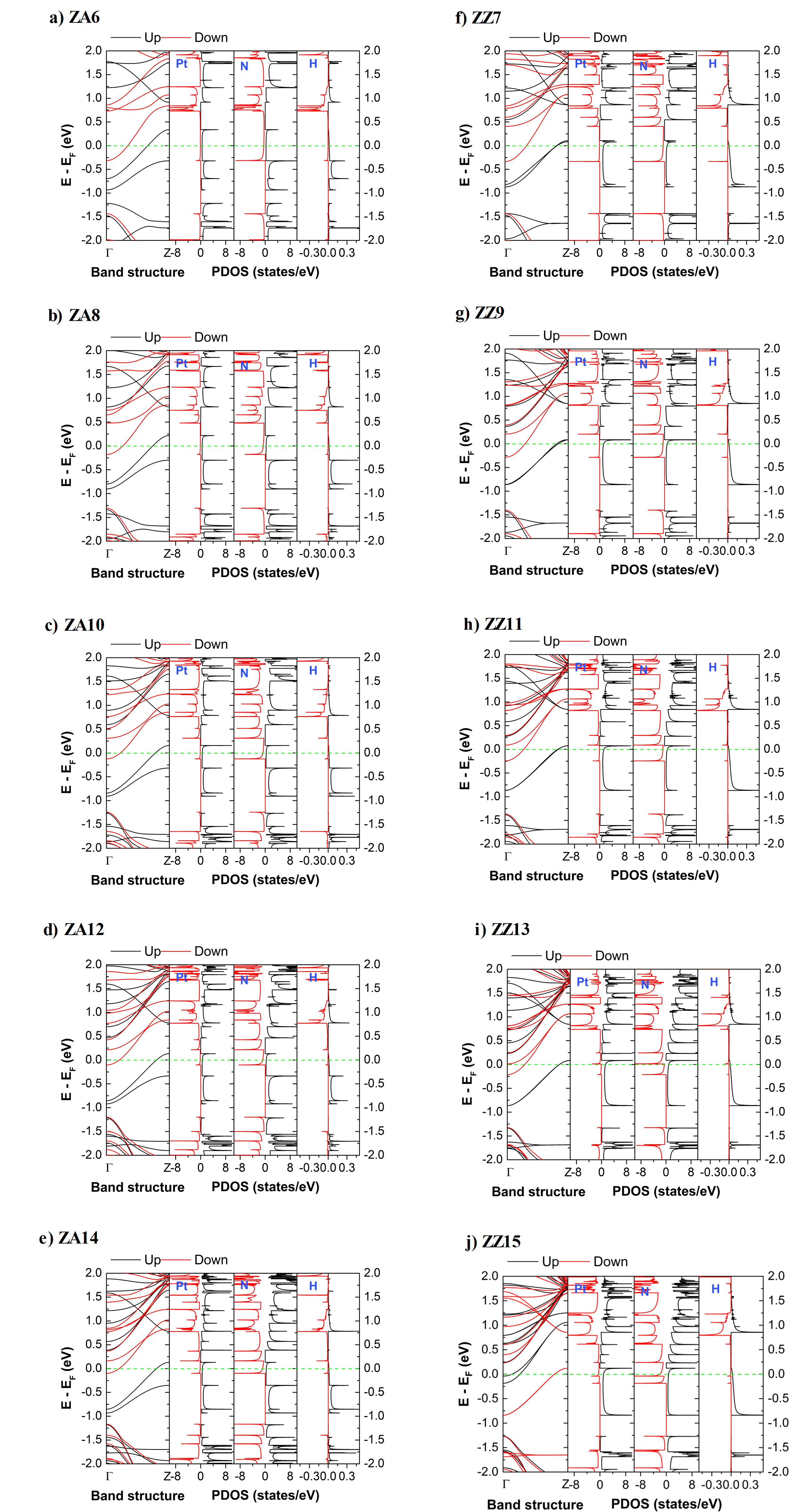}
		\caption{Spin-splitting band structure and atom-projected density of states (PDOS) for the ZA and ZZ edge shapes of the p-PtN$_2$NRs with different widths.}
		\label{Fig4} 
	\end{center} 
\end{figure*}

The analysis of the electronic properties of penta-PtN$_2$ nanoribbons with different edge types and widths is based on their electronic band structures and density of states (DOS). These features are clearly illustrated in Fig. \ref{Fig3} for the AA and SS edges and in Fig. \ref{Fig4} for the ZA and ZZ edges. In these figures, the spin-up states (black line) and spin-down states (red line) are shown, and the Fermi level is set to zero energy to distinguish between the valence and conduction states, as indicated by the dashed green line.

\begin{table}[]
	\centering
	\caption {Band gap [E$_{g}$ (eV)]/metallic, net magnetic moment [$M$($\mu_{B}$)], and magnetic-electronic characteristics of the studied structures.}
	\begin{center}
		\begin{tabular}{llll}
			\hline 
			Configurations & $E_{g}$ (eV)/metallic &  M($\mu_{B}$) & Characteristics \\ 
			\hline 
			AA5 & spin up: metallic 
			& & \\
			& spin down: metallic & 3.460 & metallic \\ 
			AA7 & spin up: 1.89 
			(indirect) & & \\
			& spin down: 1.33 (direct) & 4.002 &semiconductor \\ 
			AA9 & spin up: 1.5 (indirect)& & \\ 
			& spin down: 1.21 (direct) & 4.004 &  semiconductor \\ 
			AA11 & spin up: 1.25 (indirect)& & \\ 
			& spin down: 1.16 (direct) & 4.005 &  semiconductor \\
			AA13 & spin up: 1.41 (indirect)& & \\ 
			& spin down: 1.05 (direct) & 4.005 &  semiconductor \\
			SS7 & spin up: metallic 
			& & \\
			& spin down: metallic & 1.787 & metallic \\
			SS8 & spin up: metallic 
			& & \\
			& spin down: metallic & 0.611 & metallic \\  
			SS9 & spin up: metallic 
			& & \\
			& spin down: metallic & 0.672 & metallic \\ 
			SS10 & spin up: metallic 
			& & \\
			& spin down: metallic & 0.779 & metallic\\ 
			SS11 & spin up: metallic & & \\ 
			& spin down:1.43 (direct) & 2.001 &  haft metal \\
			\hline 
		\end{tabular}
	\end{center}
	\label{Egap}
\end{table}

\begin{table}[]
	\centering
	\caption {Band gap [E$_{g}$ (eV)]/metallic, net magnetic moment [$M$($\mu_{B}$)], and magnetic-electronic characteristics of the studied structures.}
	\hspace{1ex}
	\begin{center}
		\begin{tabular}{llll}
			\hline 
			Configurations & $E_{g}$ (eV)/metallic &  M($\mu_{B}$) & Characteristics \\ 
			\hline 
			ZA6 & spin up: metallic 
			& & \\
			& spin down: metallic & 3.325 & metallic \\
			ZA8 & spin up: metallic
			& & \\
			& spin down: metallic & 3.487 & metallic \\  
			ZA10 & spin up: metallic
			& & \\
			& spin down: metallic & 3.573 & metallic \\ 
			ZA12 & spin up: metallic 
			& & \\
			& spin down: metallic & 3.609 & metallic \\ 
			ZA14 & spin up: metallic & & \\ 
			& spin down:metallic & 3.613 & metallic \\
			ZZ7 & spin up: metallic 
			& & \\
			& spin down: metallic & 3.324 & metallic \\
			ZZ9 & spin up: metallic 
			& & \\
			& spin down: metallic & 3.368 & metallic \\  
			ZZ11 & spin up: metallic 
			& & \\
			& spin down: metallic & 3.405 & metallic \\ 
			ZZ13 & spin up: metallic 
			& & \\
			& spin down: metallic & 3.361 & metallic \\ 
			ZZ15 & spin up: metallic & & \\ 
			& spin down: metallic & 3.225 & metallic \\
			\hline 
		\end{tabular}
	\end{center}
	\label{Egap}
\end{table}

Besides, the values of bandgaps and magnetic moments are reported Table 1 and Table 2, it shows that different structures give rise to distinct electronic and magnetic properties. All penta-PtN$_2$ nanoribbons exhibit magnetic behavior and show spin polarization, since all structures possess nonzero net magnetic moments. The values of the total magnetic moment differ for each structural configuration as well as for different ribbon widths. This behavior is further confirmed by the pronounced spin splitting of the energy bands and the asymmetric peaks around the Fermi level in the partial density of states (PDOS). 
The AA-p-PtN$_2$ nanoribbon structures, including AA-7, AA-9, AA-11, and AA-13, exhibit semiconducting behavior with relatively large total magnetic moments ($\sim$4.0~$\mu_{B}$), which can be observed through the presence of an energy bandgap formed between the occupied highest valence band and lowest conduction band, as clearly evidenced in the electronic band structure and the projected density of states (PDOS). The PDOS spectra reveal the atomic contributions to the DOS peaks. The pronounced and spin-asymmetric peaks originate from the contributions of Pt, N, and H atoms, with the dominant contributions coming primarily from Pt and N.

The AA5 edge configuration and the ZA-p-PtN$_2$ and ZZ-p-PtN$_2$ nanoribbon structures exhibit metallic behavior, as the lowest conduction band and the highest valence band intersect at the Fermi level, while simultaneously possessing high magnetic moments, as reported in Table 2.

While all SS-edge penta-graphene nanoribbons are nonmagnetic semiconductors \cite{yuan2017electronic},in the case of p-PtN$_2$ nanoribbons, ribbon widths ranging from SS-7 to SS-10 give rise to metallic characteristics accompanied by magnetic moments varying between $\sim$0.6~$\mu_{B}$ and $\sim$1.8~$\mu_{B}$. Especially, the SS11 structure displayed in Fig. \ref{Fig4} (j) shows a half-metal that is ruled by existing a bandgap in spin down band is 1.43 eV (red line) and destroying a bandgap in spin up band (black line). It can determine that only the SS11 structure belong to ferromagnetic half-metal.

\begin{figure*}[h]
	\begin{center}
		\includegraphics[angle=0, width=0.9\textwidth]{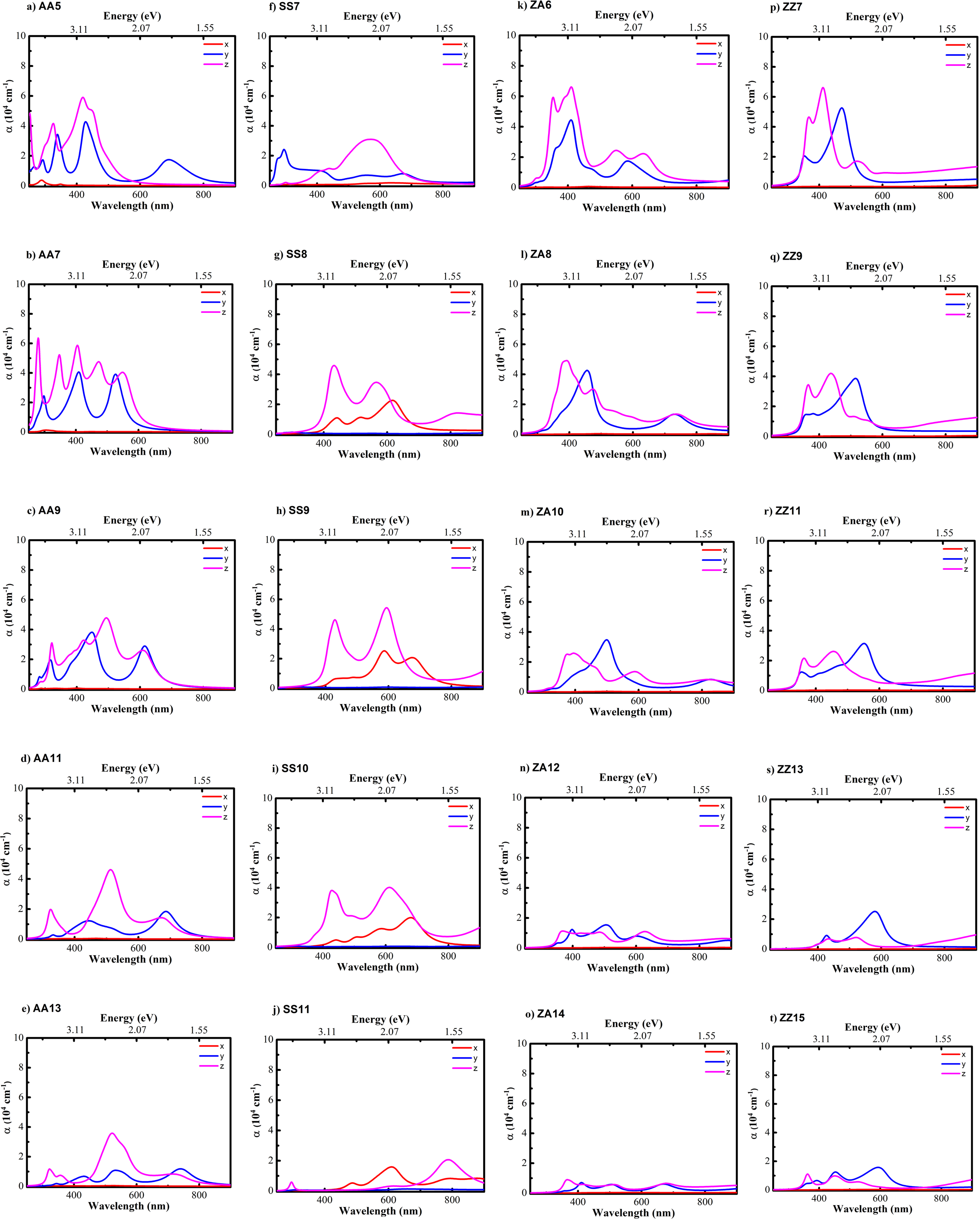}
		\caption{Spatial spin density distributions of p-PtN$_2$NRs, whereas pink and cyan isosurfaces represent for the spin up and spin down density distributions, respectively.}
		\label{Fig5} 
	\end{center} 
\end{figure*}

The investigation of spatial spin density distributions has recently attracted considerable attention.
Spin density can take either positive or negative values; by convention, the electron density associated with spin-up states is considered positive, while that corresponding to spin-down states is regarded as negative \cite{ruiz2005spin}. Accordingly, the spin density distributions of the 20 nanoribbon structures are clearly illustrated in Fig \ref{Fig5}, where the pink and cyan isosurfaces represent the spin-up and spin-down states, respectively.

As shown in Fig. \ref{Fig5}, all investigated structures are dominated by spin-up states, whereas the spin-down states are sparsely localized and make negligible contributions to the overall spin density distribution, resulting in positive net magnetic moments as presented in Table 1 and 2. The spin density is mainly localized at the edges of all structures; however, for narrower ribbons, it tends to extend into the central region.

In contrast to the SS configuration of p-PdSe$_2$, which is nonmagnetic \cite{tien2023structural}, all SS configurations of p-PtN$_2$ exhibit positive magnetic moments generated by the spin-up components as Fig .\ref{Fig5} (f-g). In particular, the SS-11 structure, which exhibits half-metallic behavior, this feature is further clarified in Fig .\ref{Fig5}(j) and Fig .\ref{Fig3}(j). A pronounced spatial asymmetry between the spin-up and spin-down density distributions is observed, with the spin-up density predominantly localized at the ribbon edges. Consistently, as shown in Fig .\ref{Fig3}(j), the spin-up bands cross the Fermi level, whereas the spin-down bands open an energy gap.

For the AA, ZA, and ZZ configurations, the spin density distributions are in good agreement with the results obtained from the previous analyses of the band structures and density of states. The spin density in these structures exhibits a symmetric distribution along both edges, following an odd–even rule in the magnetic moment behavior \cite{dang2025diverse,zheng2013novel}. In the AA configuration, the contribution to the magnetic moment originates mainly from the edge N atoms and the Pt atoms near the edges with spin-up components. As the ribbon width increases, spin-down components appear; however, their contribution is negligible, as shown in Fig. \ref{Fig5} (a-e). The ZA configuration exhibits contributions from N atoms at the armchair edge as well as both N and Pt atoms at the zigzag edge. Nevertheless, according to Fig. \ref{Fig5} (k–o), the contribution of the N atoms at the zigzag edge is the most dominant. For the ZZ configuration, both N and Pt atoms at the edge sites contribute to the magnetic moment, with the contribution from N atoms being higher, as clearly demonstrated in Fig. \ref{Fig5} (p-t).
Overall, these analyses reveal the diversity in spin arrangements and in the atomic contributions to the total magnetic moment.

\begin{figure*}[h]
	\begin{center}
		\includegraphics[angle=0, width=0.9\textwidth]{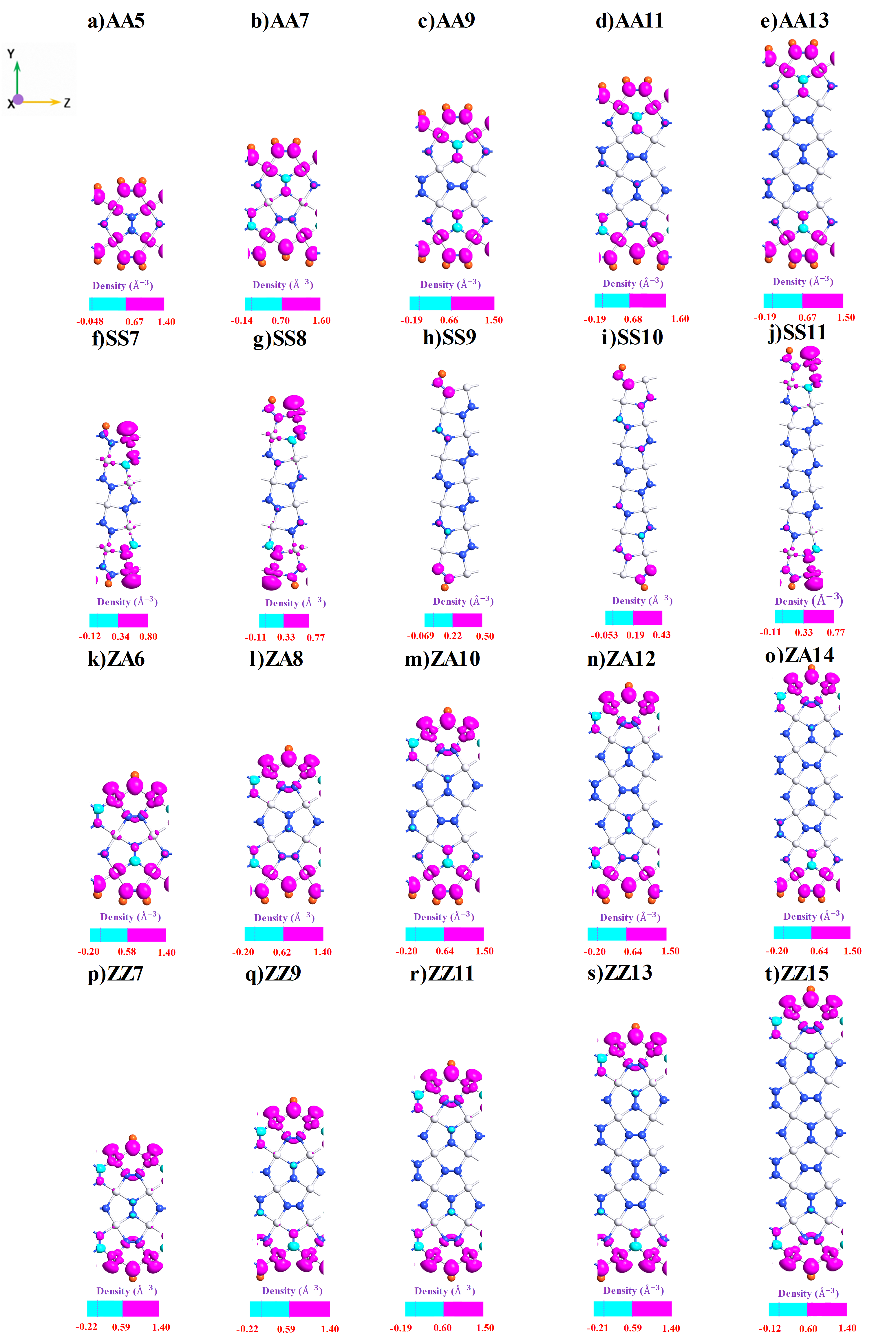}
		\caption{Optical absorption spectra of four edge types of p-PtN$_2$NRs.}
		\label{Fig6} 
	\end{center} 
\end{figure*}

Based on the calculation of the real and imaginary parts of the complex dielectric function as a function of energy, the optical properties of the p-PtN$ _{2} $ nanoribbons are presented in the Fig. \ref{Fig6}. As shown in the figure, when the width of the nanoribbons increases, the main peaks of the absorption spectra shift toward lower energy regions. The dominant peaks mainly appear along the Oy or Oz directions. Most of the investigated structures are sensitive in the visible-light region (400–600 nm).While the optical properties of the SS-edged p-NiN$ _{2} $ structures exhibit main absorption peaks along the z-axis located in the ultraviolet region with wavelengths extending up to 1000 nm \cite{dang2025diverse}, the SS-edged p-PtN$ _{2} $ structures also show dominant peaks along the z-axis but within the visible-light region. Notably, only the SS-11 configuration presents its main z-axis peak shifted toward the infrared region, with a wavelength of approximately 800 nm.

For the p-AA-PtN$ _{2} $ configurations, the dominant absorption peaks are oriented along the Oz axis, and their positions shift only toward lower-energy regions with increasing ribbon width, without altering their polarization direction. Only the AA7 structure exhibits a main peak shifted into the ultraviolet region. 
In contrast, for the p-ZA-PtN$ _{2} $ configurations, the dominant peaks are oriented along the Oz direction at small widths but switch to the Oy direction as the width increases; however, for the ZA-14 structure, the dominant peak remains along the Oz direction. 
Consistent with the optical absorption behavior of ZZ-p-NiN$ _{2} $ nanoribbons, the ZZ-p-PtN$ _{2} $ nanoribbons show a dominant absorption peak polarized along the Oz direction at approximately 500 nm for ribbon widths below 10. With further increasing width, the main absorption peak is found to shift to the Oy direction, occurring at a wavelength of about 600 nm \cite{dang2025diverse}.
Thus, the absorption spectra of p-PtN$_2$ nanoribbons can be tuned by adjusting the ribbon width and edge shape, which offers significant advantages for the development of practical applications in optoelectronic and photonic devices.

\subsection{Defect structure}

\begin{figure*}[h]
	\begin{center}
		\includegraphics[angle=0, width=0.6\textwidth]{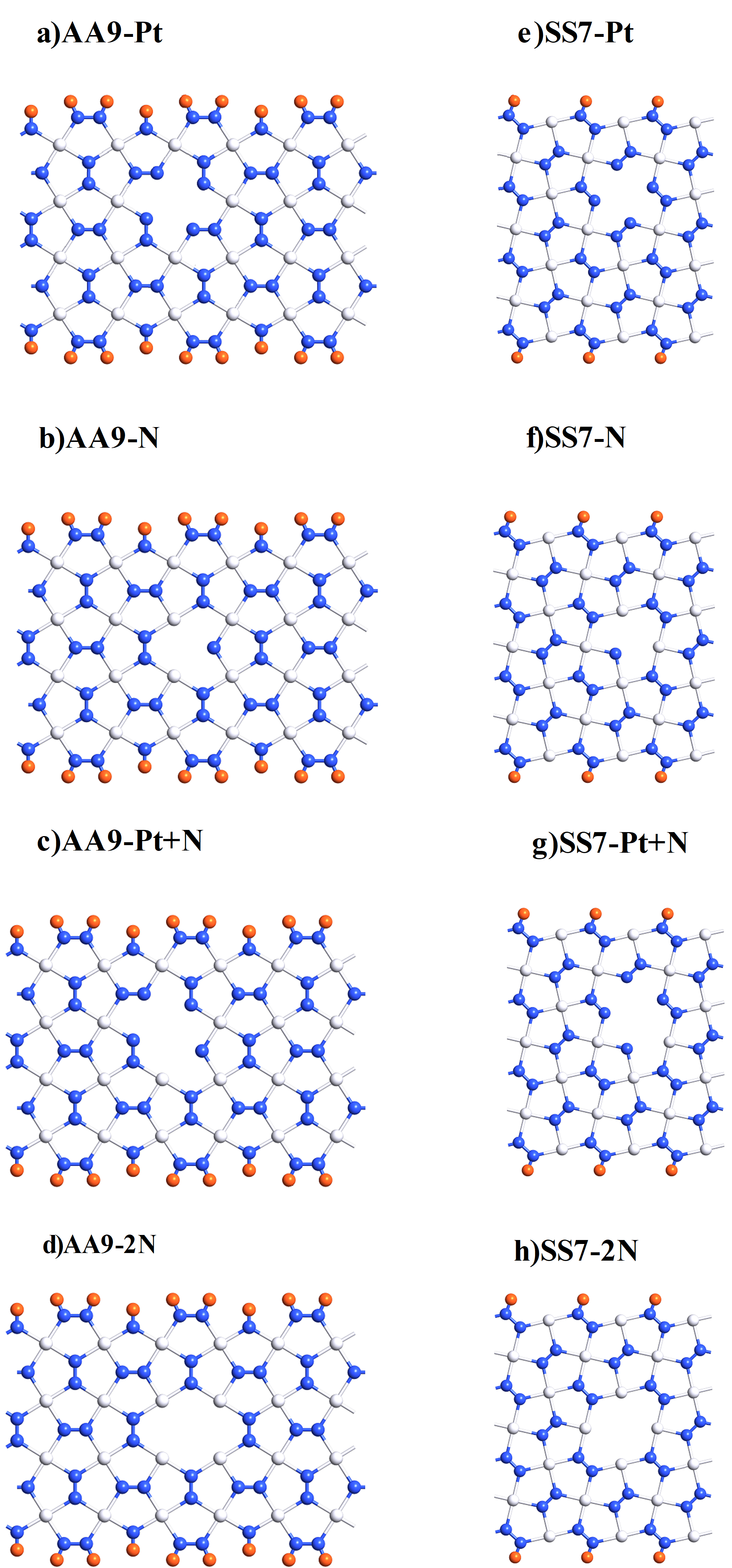}
		\caption{Defect structures of the AA9 and SS7 edges before optimization. The white, blue, and orange balls represent Pt, N and H atoms, respectively.}
		\label{Fig7} 
	\end{center} 
\end{figure*}

\begin{figure*}[h]
	\begin{center}
		\includegraphics[angle=0, width=0.6\textwidth]{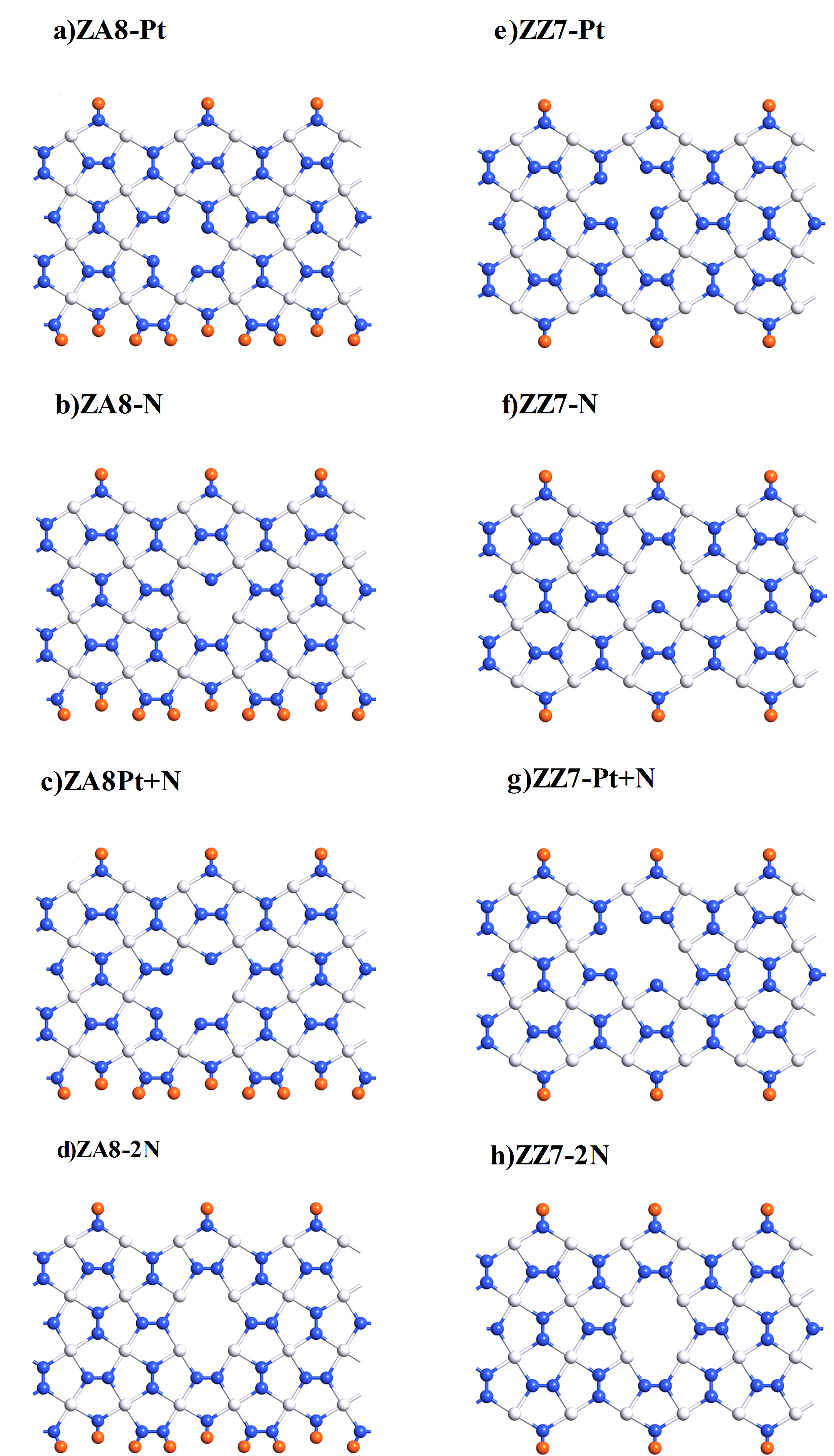}
		\caption{Defect structures of the ZA8 and ZZ7 edges before optimization. The white, blue, and orange balls represent Pt, N and H atoms, respectively.}
		\label{Fig8} 
	\end{center} 
\end{figure*}

In this section, we investigate four types of defect structures, including a single Pt vacancy (ZZ7-V$_{Pt}$), a single N vacancy (ZZ7-V$_N$), a combined Pt–N vacancy (one Pt atom and one N atom removed - ZZ7-V$_{Pt+N}$)), and a double N vacancy (ZZ7-V$_{2N}$), for the AA9, SS7, ZA8, and ZZ7 structures, as shown in Fig. \ref{Fig7} and Fig .\ref{Fig8}.

\begin{figure*}[h]
	\begin{center}
		\includegraphics[angle=0, width=0.9\textwidth]{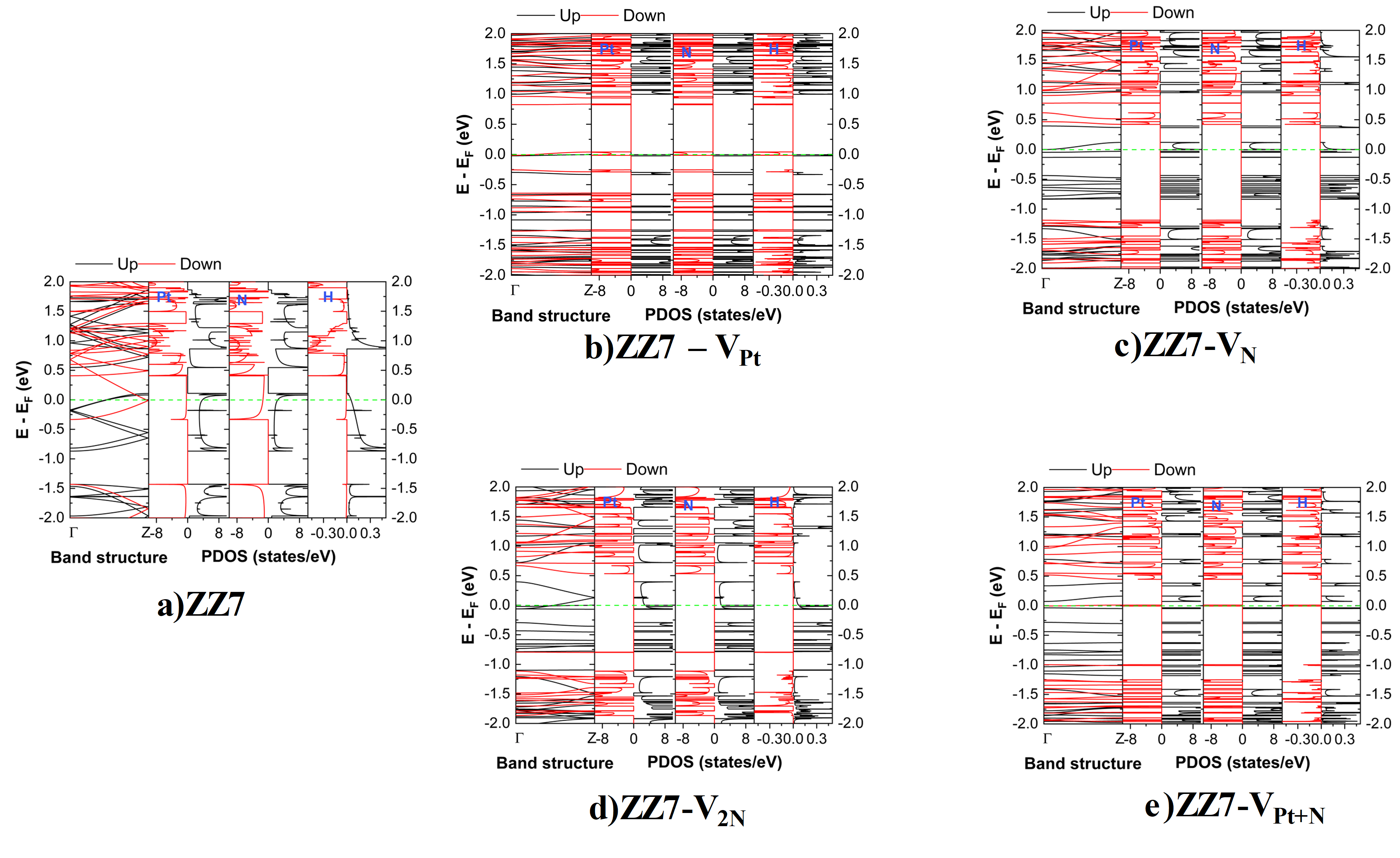}
		\caption{Spin-splitting band structure and atom-projected density of states (PDOS) for the ZZ7, ZZ7-V$_{Pt}$, ZZ7-V$_N$, ZZ7-V$_{2N}$, ZZ7-V$_{Pt+N}$ of ZZ7-p-PtN$_2$NRs structures.}
		\label{Fig9} 
	\end{center} 
\end{figure*}

The electronic structures and the contributions of Pt, N and H atoms to the density of states of the pristine ZZ7 structure, ZZ7 with a single Pt vacancy, ZZ7 with a single N vacancy, ZZ7 with two N vacancies, and ZZ7 with combined Pt and N vacancies are presented in Fig .\ref{Fig9}.
As shown in Fig .\ref{Fig9} a, the pristine ZZ7 structure exhibits metallic behavior in both the spin-up and spin-down channels. In contrast, the absence of atoms significantly alters the electronic properties of the ZZ7 structure, causing a transition from a metallic state to half-metallic.
Four types of defect structures including the ZZ7-V$_N$, ZZ7-V$_{Pt}$, ZZ7-V$_{2N}$, and ZZ7-V$_{Pt+N}$ configurations exhibit half-metallic characteristics, in which one spin channel shows semiconducting behavior while the other spin channel touches or crosses the Fermi level as Fig .\ref{Fig9} b, c, d, e.

Overall, these results demonstrate that atomic vacancies play a crucial role in tailoring the electronic structure of ZZ7 nanoribbons, enabling effective modulation of their spin polarization and band characteristics.

\begin{figure*}[h]
	\begin{center}
		\includegraphics[angle=0, width=0.9\textwidth]{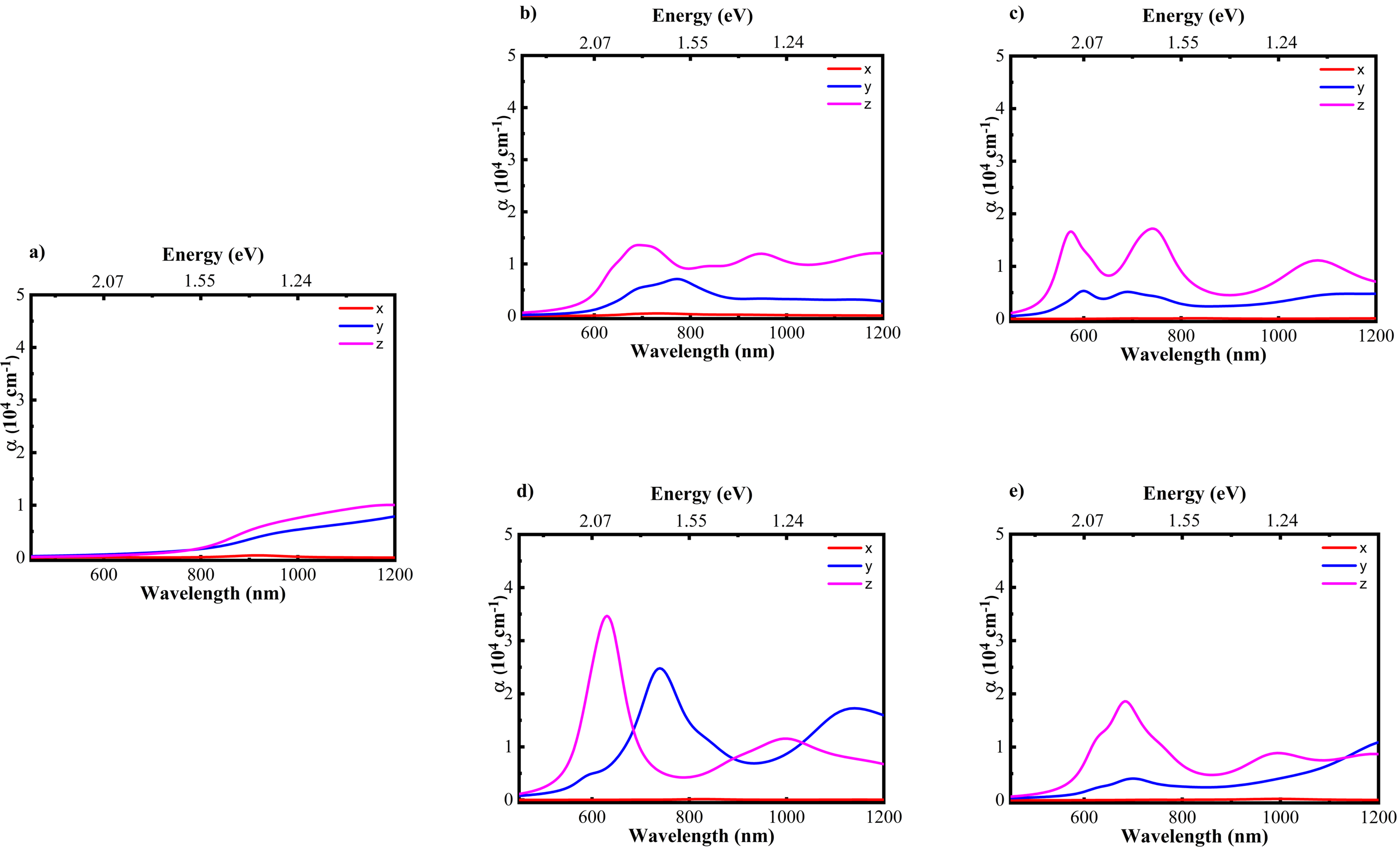}
		\caption{Optical absorption spectra of ZZ7, ZZ7-V$_{Pt}$, ZZ7-V$_N$, ZZ7-V$_{2N}$, ZZ7-V$_{Pt+N}$ of ZZ7-p-PtN$_2$NRs structures.}
		\label{Fig10} 
	\end{center} 
\end{figure*}

The optical properties of the ZZ7 structures are showed in Fig .\ref{Fig10}. The pristine structure and the defective configurations exhibit distinct optical absorption peaks polarized along the z direction. The pristine ZZ7 structure shown in Fig .\ref{Fig10} shows absorption peak appears along the z direction at a wavelength of approximately 1200 nm, which lies in the near-infrared region. 
In contrast, the ZZ7-V$_{Pt}$, ZZ7-V$_N$, ZZ7-V$_{2N}$, and ZZ7-V$_{Pt+N}$ structures exhibit notable changes in optical absorption characteristics as shown in the corresponding Fig .\ref{Fig10} b, Fig .\ref{Fig10} c, Fig .\ref{Fig10} d, Fig .\ref{Fig10} e. They exhibit optical absorption peaks in the visible light region with wavelengths. Therefore, these results clearly demonstrate that the optical properties of ZZ7 nanoribbons are highly sensitive to atomic defects. The removal of one or two atoms leads to significant increased absorption capacity in the visible range. The effects of defects may open up promising application potential in optoelectronic devices.

\section{Conclusions}
\noindent  Based on density functional theory with the highly accurate HSE06 hybrid functional, we systematically investigate the structural, electronic, optical and magnetic properties of p-PtN$_2$ nanoribbons with four representative edge terminations, namely armchair–armchair (AA), sawtooth–sawtooth (SS), zigzag–armchair (ZA), and zigzag–zigzag (ZZ), as well as their defective structures. The p-PtN$_2$ nanoribbon structures exhibit structural stability as evidenced by their significantly negative formation energies (range of -3.8 to -4.3 eV), among which the SS-edge configuration shows the highest stability compared with the other edge types. Regarding the optical absorption spectra of p-PtN$_2$ nanoribbons, they indicate that the absorption region can be flexibly tuned by varying the ribbon width and edge shape. For most structures, the main peaks predominantly appear along the Oy or Oz directions. This study demonstrates that the electronic and optical properties of p-PtN$_2$  nanoribbons  are strongly dependent on their edge configurations and ribbon widths. Defects in the nanoribbon structure also alter the electronic and optical properties compared to the pristine structure.
The flexible tunability of these properties indicates that both p-PtN$_2$  nanoribbons, and their defective configurations, are highly promising candidates for next-generation nanoelectronic materials and devices.

\section*{ACKNOWLEDGMENTS}
\noindent This research is funded by Ministry of Education and Training.

\bibliography{bib} 
\bibliographystyle{apsrev}


\end{document}